\documentclass[12pt,preprint]{aastex}

\shorttitle{The Secular Dynamics of the HD 69830 System}
\shortauthors{Ji Jianghui et al.}

\begin{document}

\title{The Secular Evolution and Dynamical Architecture of the Neptunian Triplet Planetary System HD
69830}

\author{Jianghui JI\altaffilmark{1,2,3}, Hiroshi Kinoshita\altaffilmark{4}, Lin
LIU\altaffilmark{5}, Guangyu LI\altaffilmark{1,2}}
\email{jijh@pmo.ac.cn }

\altaffiltext{1}{Purple  Mountain  Observatory , Chinese  Academy
of  Sciences ,  Nanjing  210008,China}

\altaffiltext{2}{National Astronomical Observatory, Chinese
Academy of Sciences,Beijing 100012, China}

\altaffiltext{3}{Department of Terrestrial Magnetism, Carnegie
Institute of Washington, 5241 Broad Branch Road NW, Washington, DC
20015-1305}

\altaffiltext{4}{National Astronomical Observatory,
 Mitaka, Tokyo 181-8588,Japan}

\altaffiltext{5}{Department of Astronomy,  Nanjing University,
Nanjing  210093, China}

\begin{abstract}
We perform  numerical simulations to study the secular orbital
evolution and dynamical structure in the HD 69830 planetary system
with the best-fit orbital solutions by Lovis and coworkers (2006).
In the simulations, we show that the triplet Neptunian system can
be stable at least for 2 Gyr and the stability would not be
greatly influenced even if we vary the planetary masses from
Neptune-mass to Jupiter-mass. In addition, we employ the
Laplace-Lagrange secular theory to investigate the long-term
behaviors of the system, and the outcomes demonstrate that this
theory can well describe and predict the secular orbital evolution
for three Neptune-mass planets, where the secular periods and
amplitudes in the eccentricities are in good agreement with those
of the direct numerical integrations. We first reveal that the
secular periods of the eccentricity $e_{1}$ and $e_{2}$ are
identical about 8,300 yr, i.e., $P_{1}=P_{2}\simeq 2\pi/|g_{1}-
g_{2}|$ ($g_{1,2}$ are respectively, the eigenfrequencies of the
system), while the secular variation of $e_{3}$ for the outermost
planet has a period of  $\sim18,200$ yr. Moreover, we extensively
explore the planetary configuration of three Neptune-mass
companions with one massive terrestrial planet residing in 0.07 AU
$\leq a \leq 1.20$ AU, to examine the asteroid structure in this
system. We underline that there are stable zones at least $10^{5}$
yr for low-mass terrestrial planets locating between 0.3 and 0.5
AU, and 0.8 and 1.2 AU with final eccentricities of $e < 0.20$.
Still, we also find that the secular resonance $\nu_{1}$ and
$\nu_{2}$ arising from two inner planets can excite the
eccentricities of the terrestrial bodies, and the accumulation or
depletion of the asteroid belt are also shaped by orbital
resonances of the outer planets, for example, the asteroidal gaps
at 2:1 and 3:2 mean motion resonances (MMRs) with Planet C, and
5:2 and 1:2 MMRs with Planet D. In a dynamical sense, the proper
candidate regions for the existence of the potential terrestrial
planets or Habitable Zones (HZs) are 0.35 AU $< a < $ 0.50 AU, and
0.80 AU $< a <$ 1.00 AU for relatively low eccentricities, which
makes sense to have the possible asteroidal structure in the
system.
\end{abstract}


\section{Introduction}
To date, more than 200 extrasolar planets have been discovered
about the nearby stars within 200 pc (Butler et al. 2006; The
Extrasolar Planets Encyclopaedia) mostly by the measurements of
Doppler surveys\footnote{http://www.exoplanets.org and
http://exoplanet.eu/} and transiting techniques. The increasing
numbers of the extrasolar planets are greatly attributed to the
increasing of precision of measurement. At present, the
observational precision has been achieved to $\sim 1$ ms$^{-1}$ to
$3$ ms$^{-1}$, whereas most observations currently have lower
precision. The improvement of the observations will indeed induce
the substantial discovery: (1)plentiful multiple systems, which
$\sim 20$ multiple system involved in orbital resonance and
secular interactions now have been detected; (2) much more
low-mass companions around main-sequence stars (from Neptune-mass
to Earth-mass, so-called super-Earths), e.g., 55 Cancri (McArthur
et al. 2004), GJ 876 (Rivera et al. 2005), HD 160691 (Santos et
al. 2004; Gozdziewski et al. 2006);(3)a true Solar System analog,
bearing several terrestrial planets, the asteroidal structure and
the dynamical environment of potentially terrestrial planets in
the Habitable Zones (HZs) (Kasting et al. 1993) for the
development of life; (4)the census of the systems, and the
diversity of the planetary systems may provide abundant clues for
theorists to more accurately model the planetary formation (Ida \&
Lin 2004; Boss 2006).

Lovis et al. (2006) (hereafter Paper I) recently reported the
discovery of an interesting system of three Neptune-mass planets
orbiting about HD 69830 through high precision measurements with
the HARPS spectrograph at La Silla, Chile. The nearby star HD
69830 is of spectral type K0V with an estimated mass of $0.86 \pm
0.03 M_{\odot}$ and a total luminosity of $0.60 \pm 0.03
L_{\odot}$ (Paper I), about 12.6 pc away from the Sun. In
addition, Beichman et al. (2005) announced the detection of a
large excess owing to hot grains of crystalline silicates orbiting
the star HD 69830 and inferred that there could be probably a
massive asteroid within 1 AU. Subsequently, Alibert et al. (2006)
and Paper I performed lots of calculations to simulate the system
and revealed that the innermost planet may possess a rocky core
surrounded by a tiny gaseous envelope and probably formed inside
the ice line in the beginning, whereas the two outer companions
formed outside the ice line from a rocky embryo and then accreted
the water and gas onto the envelope in the sequent formation
process. Hence, it is important for one to understand the
dynamical structure in the final assemblage of the planetary
system (Asghari et al. 2004; Ji et al. 2005; Jiang \& Yeh 2006),
and to investigate suitable HZs for life-bearing terrestrial
planets (David et al. 2003; Fatuzzo et al. 2006; Jones et al.
2005; Haghighipour 2006) advancing the space missions (such as
\textit{CoRot, Kepler} and \textit{TPF}) aiming at detecting them,
thus one of our goals is to focus on this issue (\S2.3).

On the other hand, from a dynamical viewpoint, one may be
concerned about the long-term stability of the system. In the
present work, we further perform a large number of simulations of
this system to examine the stability (\S2.1) and to understand the
secular evolution by means of the secular theory (\S2.2). Finally,
we summary the main results and present a concise discussion in
\S3.

\section{Dynamical Analysis}
In this paper, we adopt the orbital parameters of the HD 69830
system provided by Paper I. In the simulations, we always take the
stellar mass and the minimum planetary masses from Table 1, except
where noted. To be specific, the mass of the host star is
$0.86M_{\odot}$, and the masses of three Neptunian planets are
respectively, $10.2M_{\oplus}$, $11.8M_{\oplus}$ and
$18.1M_{\oplus}$, where $\sin i=1$.  We utilize N-body codes (Ji,
Li \& Liu 2002) to perform numerical simulations by using RKF7(8)
and symplectic integrators (Wisdom \& Holman 1991) for this
system. In the numerical study, the adopted time stepsize is
usually $\sim$ 2\% - 5\% of the orbital period of the innermost
planet. In addition, the numerical errors were effectively
controlled over the integration timescale, and the total energy is
generally conserved to $10^{-6}$  for the integrations. The
typical timescale of simulations of the three-planet system is
from 100 Myr to 2000 Myr.

\subsection{The Stability of the  HD 69830 Planetary System}
To examine the secular stability of this system, we numerically
integrated the three-planet system on a timescale of 2 Gyr in
terms of the initial values listed in Table 1.  In Figure 1, a
snapshot of the secular orbital evolution of three planets is
illustrated, where $Q_{i}=a_{i}(1+e_{i})$, $q_{i}=a_{i}(1-e_{i})$
(the subscipt $i=1,2,3$, individually, denoting Planet B, C and D)
are, respectively, the apoapsis and periapsis distances. In the
secular dynamics, the semi-major axis $a_{1}$ remains unchanged to
be 0.0785 AU for 2 Gyr, whilst $a_{2}$ and $a_{3}$ slightly
undergo librations about 0.186 and 0.630 AU with small amplitudes
over the same timescale. The variations of eccentricities during
long-term evolution are followed, where $0.05 < e_{1}< 0.20$, $0.0
< e_{2}< 0.15$ and $0.069 < e_{3}< 0.078$. In Fig.1, we note that
the time behaviors of $Q_{i}$ and $q_{i}$ again show regular
motions of bounded orbits for three planets, indicating their
orbits are well separated in the secular evolution due to smaller
mutual interactions. In the numerical study, we find the system
can be dynamically stable and last at least for 2 Gyr, and our
simulations do extend and confirm the numerical exploration of 1
Gyr by Paper I.

To explore the stability of HD 69830 with respect to the
variations of the planetary masses, we first fix $\sin i$ in
increment of 0.1 from 0.1 to 0.9. In the additional numerical
experiments, we simply alter the masses but keep all orbital
parameters (Table 1), again restart new runs of integration of
three-planet system for 100 - 1000 Myr with the rescaled masses.
As a result, we find the system could remain definitely stable for
the above investigated timescale with slight vibrations in
semi-major axes and eccentricities for all planets. However, Paper
I also pointed out that the system can even survive for the
planets bearing the Jupiter-masses. To verify this, we further let
$i=1.0^{\circ}, 0.8^{\circ}, 0.5^{\circ}$, then the planetary
masses are in the Jupiter-mass range, and the outcomes for new
integrations of 100 Myr show that there is no chaotic behavior
occurring in this system. This may also imply that the stability
of the HD 69830 system is not so sensitive to the planetary
masses.

Next, we come to the qualitative stability criteria for this
system. Gladman (1993) analytically attained the minimum
separation $\Delta$ between two planets of masses $m_{1}$ and
$m_{2}$ moving on initially circular orbits required for
stability,
\begin{equation}
\label{eq311}
\Delta \simeq 2\sqrt 3 R_{H},
\end{equation}
where the mutual Hill radius  $R_{H}$ is defined by
\begin{equation}
\label{eq312}
R_{H}={\left(\frac{m_{1} +
m_{2}}{3m_{c}}\right)}^{1/3} \left({\frac{a_{1} +
a_{2}}{2}}\right).
\end{equation}
With the orbital data in Table 1, we then approximately estimate
$\Delta_{12} \simeq 0.0135$ AU, for the couple of Planet B and C;
similarly, $\Delta_{23} \simeq 0.0460$ AU,  for the pair of Planet
C and D. Note this criteria are good guidelines for stability but
are not definitive constraints for nearly-circular orbits.
However, Chambers et al. (1996) presented a more general stability
criterion that shows the initial orbital separation should be no
less than about 10 $R_{H}$ for multi-planet systems. For the
Neptune-mass planets, the actual orbital separations are $\sim 10$
times of the analytical values that the stability requires, while
the mutual distances between two orbits can amount to be $\sim$ 10
$R_{H}$, even though these planets possess Jupiter masses.
Therefore, from both the numerical and analytical views, we can
safely conclude that the HD 69830 system is dynamically stable for
the lifetime of the star.

\subsection{Secular Dynamics}
Considering a system of $N$ massive bodies of mass $m_{i}$ $(i=1,
N)$ moving about the central star of mass $m_{c}$ under their
mutual gravitational forces and the attraction of the central
body, if $m_{i}\ll m_{c}$, and these planets are with the orbits
of small eccentricities and low inclinations, then the long-term
evolution of the eccentricities and inclinations of all planets
can be analytically described by classical Laplace-Lagrange
(hereafter L-L) secular theory, provided that the mean motion
resonance amongst them is absent.  Many authors have adopted the
secular theory to study the orbital evolution of the planets in
the systems, e.g., Ji et al. (2003) applied this theory to explore
the dynamical evolution of three giant planets in 55 Cancri; more
recently, Adams \& Laughlin (2006; and references therein) further
investigated the dynamical interactions in the multiple planetary
system by using secular theory and showed that the orbital
eccentricities in these systems alter over secular time scales, of
typical variations over $10^{3} - 10^{5}$ years.

As for the HD 69830 system, there are three Neptune-mass planets
orbiting the host star, assuming all planets are coplanar.
Furthermore, in Table 1 we notice that the orbital eccentricities
are smaller than about 0.10, hence one may realize that the L-L
secular theory can be applied to this system. As usual, the L-L
secular theory tells us the story of the precession of the
longitudes of periastron and that of the longitudes of node, and
the variations of the amplitudes of the eccentricities and
inclinations. Herein, we simply address the issue of the secular
evolution of the eccentricities. Bearing this in mind, the L-L
theory is an approximation to the actual motions, because it is
valid only to the second order in the eccentricities and
inclinations and the first order in the planetary masses, however,
it may reveal substantial clues for long-term evolution of the
system. Subsequently, we will use the analytical results from L-L
theory to compare with the numerical outcomes from direct
integrations.

The linear L-L secular solutions for the eccentricities (the
eccentricity eigenvectors, see Murray \& Dermott 1999, MD99) can
be written :
\begin{equation}
\label{eq1}
{
e_{j}\sin\varpi_{j} = \sum_{i=1}^{3}e_{ji}\sin(g_{i}t+\beta_{i})
}
\end{equation}
\begin{equation}
\label{eq2}
{
e_{j}\cos\varpi_{j} = \sum_{i=1}^{3}e_{ji}\cos(g_{i}t+\beta_{i})
}
\end{equation}
where $e_{ji}$, $g_{i}$ and $\beta_{i}$ ($j=1,2,3$, respectively
for Companions B, C and D; $i=1,2,3$) are respectively, the
magnitudes of eccentricity eigenvectors, the eigenfrequencies and
the phases, which can be calculated from the initial orbital
solutions given in Table 1. In Table 2 are listed the magnitudes
of the eccentricity eigenvectors $e_{ji}$ evaluated from the
secular theory by considering all three planets. Firstly, we
obtain three eigenfrequencies $g_{1}=236^{"}.02$ yr$^{-1}$,
$g_{2}=79^{"}.76$ yr$^{-1}$ and $g_{3}=8^{"}.49$ yr$^{-1}$ (see
Table 2) for the HD 69830 system, and these values are related to
the precession of the periastron of three Neptune-mass planets
with periods of 5,491 yr, 16,248 yr and 152,650 yr, which are
essentially in agreement with the results of Paper I. In addition,
the phases are $\beta_{1}=12.12^{\circ}$,
$\beta_{2}=244.13^{\circ}$ and $\beta_{3}=224.56^{\circ}$, and now
all constants in the right-side of Eqs. (\ref{eq1}) and
(\ref{eq2}) are well determined, hence we can derive the
analytical expressions of the eccentricities $e_{j}$ for Planet
$j$ at any time $t$, and its squares are given by
\begin{equation}
\label{eq3} { e_{j}^{2} = (\sum_{i=1}^{3}
e_{ji}\sin(g_{i}t+\beta_{i}))^{2} + (\sum_{i=1}^{3}
e_{ji}\cos(g_{i}t+\beta_{i}))^{2}}
\end{equation}

Secondly, we estimate the period of $e_{1}$ in the secular orbital
evolution using the L-L theory, from Table 2, we note that the
magnitudes of the eccentricity vectors $e_{11}, e_{12} \gg
e_{13}$, and it is not difficult for one to understand that the
mutual interactions of two inner planets are much greater than
that of the distant outermost planet. Hence, we have
\begin{equation}
\label{eq4} {e_{1}^{2} \simeq e_{11}^2 + e_{12}^2 +
2e_{11}e_{12}\cos((g_{1}-g_{2})t + (\beta_{1}-\beta_{2})). }
\end{equation}
Then, the secular period of $e_{1}$ given by analytical method is
$P_{1}\simeq 2\pi/(g_{1}- g_{2})$, about 8,300 yr. Similarly, the
secular period of $e_{2}$ can be also derived, where $P_{2}\simeq
2\pi/|(g_{2}- g_{1})|$, apparently, is equal to $P_{1}$. Now, we
compare the analytical predictions with the numerical
integrations. In Fig. 2, a snapshot is illustrated to perform the
comparisons, while it can reflect the situation of total timescale
of 2 Gyr. In this figure, the \textit{dash dot line} denotes the
variational eccentricities of numerical results, while the
\textit{solid line} represents those from secular theory.  For the
upper and middle panels, $P_{1}$ and $P_{2}$ are clearly evident
and also in accordance with the corresponding long-term periods
for eccentricities in the numerical results. In the meantime, the
librating amplitudes for the eccentricities $e_{1}$, $e_{2}$ from
the L-L theory (\textit{solid line}), where $(e_{1min},
e_{1max})\simeq (|e_{11} - e_{12}|, |e_{11} + e_{12}|)=(0.063,
0.193)$, then $e_{1}$ is in the range [0.05, 0.20] and $e_{2}$ in
[0.0, 0.15], are well coupled with those of the numerical results
(\textit{dash dot line}) by long-term integrations. Moreover, we
also notice that $e_{1}$ experiences to be maximum while $e_{2}$
goes down minimum in the orbital evolution (see Fig.2), and
\textit{vice versa}, still this is because of the conversation of
total angular momentum for the system. As aforementioned, the
mutual perturbations between two inner planets are much more
important than their interactions with the outer planet (see Table
2). On the other hand, in the secular evolution over 2 Gyr,
$a_{1}$ and $a_{2}$ almost keep unchanged (Fig. 1), by ignoring
Planet D, thus we have the total angular momentum
$c_{1}\sqrt{1-e_{1}^2}+c_{2}\sqrt{1-e_{2}^2} =$const ($c_{1,2}$
are the combination constants of the planetary masses and
semi-major axes), hence we obtain the maximum-minimum correlation
of $e_{1}$ and $e_{2}$, \textit{de facto}, such orbital evolution
may again imply two coupled secular periods of the eccentricity in
this system.

As to Planet D, the contributions of the innermost planet (HD
69830 b) is much less than that of the intermediate companion (HD
69830 c), indicating the mutual interaction from Planet C and D is
dominant, then we neglect the relevant eccentricity items arising
from Planet B in (\ref{eq3}), and the secular period of the
eccentricity of the outermost planet is achieved, where $P_{3}
\simeq 2\pi/(g_{2} - g_{3})$ about 18,200 yr. For the bottom panel
in Fig. 2, one can observe the long-term eccentricity periods of
$P_{3}$ of the secular theory and numerical simulations coincide
with each other. Moreover, the secular theory presents the scope
of $e_{3}$ in [0.069, 0.078], which also confirms the direct
numerical simulations. Lastly, we reach the conclusion that the
outcomes from L-L theory\footnote{Adams \& Laughlin (2006) studied
the secular theory (MD99) by adding leading-order correction for
the general relativity (GR) and showed that GR may act to either
stabilize or destabilize the system depending on the specific
architecture of the planetary system. To examine whether GR may
play a role in the stability of this system, we again numerically
perform new runs by modifying the equations of motion including
the GR corrections. The outcomes show that the system still retain
stable in the presence of GR.} are in accordance with the
numerical results.

\subsection{Dynamical Architecture and Potential HZs}
Modern observations by \textit{Spitzer} and \textit{HST} clearly
exhibit that the circumstellar debris disks (e.g., AU Mic and
$\beta$ Pic) are quite common in the early planetary formation
(for a recent review see Werner et al. 2006), and Beichman et al.
(2006) show that there are $13\pm3\%$ for Kuiper Belt analogs
around mature main sequence stars by \textit{Spitzer} programs,
further they point out that the existence of debris disks is
extreme important to the resulting detection of individual planets
and related to the formation and evolution of planetary systems.
As mentioned previously, Beichman et al. (2005) also provide a
clear evidence of the presence of the disk in HD 69830, and
subsequently Paper I produce the best-fit orbital solutions of
three Neptune-mass planets with well-separated nearly-circular
orbits, and meanwhile the present study confirms the secular
orbital stability in this system, which may imply that the HD
69830 system is likely to be an analog of the Solar System, then
to possess other terrestrial planets, and to be made of a
potential asteroidal belt structure or Kuiper belt analog in the
system. Hence, it deserves to make a detailed investigation from a
numerical perspective.

To further investigate the dynamical structure and potential HZs
in this system, we performed additional simulations with the
planetary configuration of coplanar orbits of three Neptune-mass
companions with one massive Earth-like planet. In this series of
runs, the mass of the assumed terrestrial planet ranges from 0.01
$M_{\oplus}$ to 1.0 $M_{\oplus}$. And the initial orbital
parameters are as follows: the numerical investigations were
carried out in [$a, e$] parameter space by direct integrations,
and for a uniform grid of 0.01 AU in semi-major axis (0.07 AU
$\leq a \leq 1.20$ AU) and 0.01 in eccentricity ($0.0 \leq e \leq
0.2$), the inclinations are $0^{\circ} < I < 5^{\circ}$, and the
remainder angles are randomly distributed between $0^{\circ}$ and
$360^{\circ}$ for each orbit, then the low-mass bodies were
numerically integrated with three Neptune-mass planets in the HD
69830 system. In total, about 2400 simulations were exhaustively
run for typical integration time spans from $10^{5}$ to $10^{6}$
yr (about $10^{6}$ - $10^{7}$ times of the orbital period of the
innermost planet). Our main results now follow.

In Figure 3, are shown the contours of the surviving time for
Earth-like planets (Fig. 3a) and the status of their final
eccentricities (Fig. 3b) for the integration over $10^{5}$ yr,
where the horizontal and vertical axes are the initial $a$ and
$e$. Fig.3a displays that there are stable zones for the low-mass
planets in the regime between 0.3 and 0.5 AU, and 0.8 and 1.2 AU
with final  eccentricities of $e < 0.20$. Obviously, there exist
unstable zones for the nearby orbits of three planets, where the
planetary embryos there have short dynamical surviving time
(usually less than $10^{3}$ yr) and the eccentricities can be
quickly pumped up to a high value $\sim 0.9$ (see Fig. 3b), where
the evolution is not so sensitive to the initial masses. In fact,
these adjacent bodies are in association with many of the mean
motion resonances of the Neptunian planets and the overlapping
resonance mechanism (MD99) can reveal their chaotic behaviors of
being ejected from the system in short dynamical lifetime,
furthermore most of their orbits are within $3R_{H}$ sphere of the
Neptune-mass planets, and others are involved in the secular
resonance with two inner companions.

Analogous to our Solar system, if consider the middle Planet C as
the counterpart as Jupiter, we will have the regions of mean
motion resonances: 2:1 (0.117 AU), 3:2 (0.142 AU), 3:1 (0.089 AU)
and 5:2 (0.101 AU), 2:3 (0.244 AU), and in Fig. 3 we notice there
indeed exist the apparent asteroidal gaps about or within above
MMRs (e.g., 3:1 and 5:2 MMRs), while in the region between 0.10 AU
and 0.14 AU for $e< 0.10$, there are stable islands that the
planetary embryos can last at least $10^{5}$ yr. In addition, for
Planet D, most of the terrestrial planets in 0.50 AU $< a < $ 0.80
AU are chaotic and their eccentricities are excited to moderate
even high values, the characterized MMRs with respect to the
accumulation or depletion of the asteroid belt are 3:2 (0.481 AU),
2:1 (0.397 AU), 5:2 (0.342 AU), 4:3 (0.520 AU), 1:1 (0.630 AU),
2:3 (0.826 AU), 1:2 (1.000 AU), and our results enrich those of
Paper I for massless bodies over two consecutive 1000-year
intervals, showing a broader stable region beyond 0.80 AU. Note
that there exist stable Trojan terrestrial bodies in a narrow
stripe about 0.630 AU, involved in 1:1 MMR with Planet D, and they
can survive at least $10^{6}$ yr with  resulting small
eccentricities in the extended integrations\footnote{It would be
likely that a fraction of the stable orbits become unstable over
even longer timescale, however these results do reveal potential
asteroid structure in this system. In a statistical sense, David
et al. (2003) investigated the stability of the Earth-like planets
in the binary systems and showed that the ejection simulations are
chaotic in the studied parameter space; in addition, Fatuzzo et
al. (2006) further focus on the characterization of the
distribution of survival times for Earth-like planets orbiting a
solar-type star with a stellar companion and present meaningful
calculations on the width of the distributions of survival time
across parameter space, and all the studies will provide helpful
information of detecting such low-mass terrestrial planets in
future survey. }. The stable Trojan configurations may possibly
take place in the extrasolar planetary systems, for example, Ji et
al. (2005) explored such Trojan planets moving about 47 Uma, and
Gozdziewski \& Konacki (2006) also argued that there may exist
Trojan pairs configurations in the HD 128311 and HD 82943 systems.
Recently, Ford \& Gaudi (2006) develop  a novel method of
detecting Trojan companions to transiting close-in extrasolar
planets and argue that the terrestrial-mass Trojans may be
detectable with present ground-based observatories, whereas the
terrestrial Trojan planets with low eccentricity orbits about 1 AU
would be probably a fascinating habitatable environment for other
life-bearing worlds.

Malhotra (2004) and Ji et al. (2005) showed that the secular
resonances may possibly exist in the general planetary systems,
and play a part in the dynamical distributions of small bodies,
then shape the asteroidal architecture as well as the orbital
resonances. In this system, if consider a planetary embryo with a
mass of $0.01 M_{\oplus}$ (in lunar mass), the related location
for $\nu_{1}$ secular resonance originating from Planet B is $\sim
0.321$ AU, where two eigenfrequencies for the terrestrial body and
innermost planet are, respectively, $232^{"}.33$ yr$^{-1}$ and
$235^{"}.94$ yr$^{-1}$, which are obtained by the L-L secular
theory, as the Earth-like planet bears the finite mass and then it
may modify the values of the \textit{in situ} eigenfrequencies
$g_{i}$ $(i=1,2,3)$ of three Neptunian planets, thus
$\nu_{i}\simeq g_{i}$, which also indicates that both planets
almost share common secular apsidal precession rates in their
motion. Fig. 3b exhibits that the eccentricity of the
planetesimals could be excited by $\nu_{1}$ about 0.32 AU, and
meanwhile the 5:2 MMR about 0.34 AU (for Planet D) drives the
eccentricities to modest values, which may imply that the
combination effects of the secular resonance and MMR can act as a
hybrid dynamical mechanism of clearing of the small bodies in the
disk. Similarly, $\nu_{2}$ resonance with respect to the
intermediate companion (Planet C) occurs for a lunar-size planet
about $\sim 0.962$ AU at an eigenfrequency of $76^{"}.72$
yr$^{-1}$, and the $\nu_{2}$ region is also close to 1:2 resonance
with the outermost Planet D, where the eccentricities for low-mass
bodies about 1 AU can be stimulated to moderate values in the
extended time integrations of $10^{6}$ yr. Our results are in
accordance with Nagasawa et al. (2005), who pointed out that the
secular resonances can pass through the terrestrial region from
outside to inside to excite the eccentricities of the separated
planetary embryos. Note that $\nu_{3}$ arises from Planet D with
an eigenfrequency of $8^{"}.62$ yr$^{-1}$, which may be at work
$\sim 1.488$ AU. In a word, the numerical study may suggest that
there exist asteroid structure of the HD 69830 system.

\section{Summary and Discussions}
In this work, we have studied the secular dynamics and dynamical
structure of HD 69830. We now summarize the main results as
follows:

(1)In the simulations, we show that the triplet Neptunian system
can be stable at least for 2 Gyr and that the stability would not
be greatly influenced when we shift the planetary masses from
Neptune-mass to Jupiter-mass. The L-L theory can well describe and
predict the secular orbital evolution for three Neptune-mass
planets, where both the secular periods and librating amplitudes
in the eccentricities are well coincidence with those obtained
from direct integrations. We first reveal that the secular periods
of the eccentricity $e_{1}$ and $e_{2}$ are identical, about 8,300
yr. Account for the nature of near-circular well-spaced orbits,
the HD 69830 system may be a close analog of the Solar System.
Nevertheless, as well-known in our Solar System, Jupiter can play
a major role as a dynamical shield to bounce and throw away most
of small bodies at times and further to prevent many asteroids
from colliding into the Earth (or other inner terrestrial planets)
over the evolution. However, if all planets of HD 69830 are in
Neptune-mass, such protection mechanism for the potential
terrestrial planets would be weaker and larger masses would be
helpful in building a harmonic habitable environment.

(2)Moreover, we extensively investigate the planetary
configuration of three Neptune-mass companions with one massive
terrestrial planet in the region of 0.07 AU $\leq a \leq 1.20$ AU
to examine the existence of the potential Earth-like planets and
further to study the asteroid structure or HZ in this system. We
show that there are stable zones at least $10^{5}$ yr for the
low-mass terrestrial planets locating between 0.3 and 0.5 AU, and
0.8 and 1.2 AU with final eccentricities of $e < 0.20$.
Furthermore, we also find that the secular resonance $\nu_{1}$ and
$\nu_{2}$ arising from two inner planets can excite the
eccentricities of the terrestrial bodies, and the accumulation or
depletion of the asteroid belt are also shaped by orbital
resonances of the outer planets, for example, the asteroidal gaps
of 2:1 and 3:2 MMRs with Planet C, and 5:2 and 1:2 resonances with
Planet D. On the other hand, the stellar luminosity of HD 69830 is
lower than that of the Sun, thus the HZ should shift inwards
compared to our Solar System. However, Kuchner (2003) pointed out
that the habitable regime is considered to be centered at
$a_{p}\geq$ 1.03 AU $(L/L_{\odot})^{1/2}$, such that a water
planet can exist in equilibrium with stellar radiation, where
$a_{p}$ is the radius of the planet's orbit and $L$ is the
luminosity of the host star, then $a_{p}$ is approximate to $0.80$
AU for HD 69830. Therefore, in a dynamical consideration, the
proper candidate regions for the existence of the potential
terrestrial planets or HZs are 0.35 AU $< a < $ 0.50 AU, and 0.80
AU $< a <$ 1.00 AU for relatively low eccentricities. Hopefully,
the future space-based observations, e.g., \textit{CoRot, Kelper}
and \textit{TPF} will definitely present a handful of samples
belonging to the category of the terrestrial bodies. In final, we
may summarize that the HD 69830 system can possess the asteroidal
architecture resembling to the Solar System and both the mean
motion resonance (MMR) and secular resonances (Malhotra 2004; Ji
et al. 2005; Nagasawa et al. 2005) will work together to influence
on the distribution of the small bodies in the planetary system.
However, the formation of this system is still a mystery and needs
further investigation in forthcoming study.

\acknowledgments{We would like to thank the anonymous referee for
a prompt report and valuable comments and suggestions that help to
improve the contents. J.H.J. is much grateful to Paul Butler and
Megan Falla for their hospitality during his stay of visit at DTM
and also appreciates to the astronomy group therein: Sean Solomon,
Alan Boss, John Chambers, Sara Seager, Alycia Weinberger, Sandy
Keiser, Vera Rubin, Mercedes Lopez-Morales, and Kevin Wang, Terry
Stahl, Janice Dunlap, Alicia Case and many others for their
kindness, assistance and helpful discussions. We are also thankful
to Zhou Q.L. for the assistance of computer utilization. Part of
the computations were carried out on high performance workstations
at Laboratory of Astronomical Data Analysis and Computational
Physics of Nanjing University. This work is financially supported
by the National Natural Science Foundations of China (Grants
10573040, 10673006, 10203005, 10233020) and the Foundation of
Minor Planets of Purple Mountain Observatory.}

\clearpage
\begin{deluxetable}{llll}
\tablewidth{0pt}
\tablecaption{The orbital parameters of HD 69830 planetary system}
\tablehead{\colhead{Parameter} & \colhead{Companion B} &\colhead{Companion C} & \colhead{Companion D}}
\startdata
$M$sin$i$($M_{\oplus}$)\tablenotemark{a}
                                 & 10.2      & 11.8      & 18.1         \\
Orbital period $P$(days)         & 8.667     & 31.56     & 197          \\
$a$(AU)                          & 0.0785    & 0.186     & 0.630        \\
Eccentricity $e$                 & 0.10      & 0.13      & 0.07         \\
Longitude of Peri. $\varpi$(deg) & 340       & 221       & 224          \\
Epoch of Peri.(JD)               & 2453496.8 & 2453469.6 & 2453358
\enddata
\tablenotetext{a}{The parameters are taken from Paper I. The mass
of the star is $0.86M_{\odot}$.}
\end{deluxetable}

\begin{deluxetable}{lrrr}
\tablewidth{0pt} \tablecaption{The values of $e_{ji}$
($\times10^8$)($j=1,2,3$, respectively for Companions B, C and D;
$i=1,2,3$) and the eigenfrequencies $g_{j}$ are given from secular
Laplace-Lagrange theory for HD 69830 planetary system, and the
unit for $g_{j}$ is yr$^{-1}$}

\tablehead{\colhead{$e_{ji}$} &\colhead{1} &\colhead{2}
&\colhead{3\tablenotemark{b}}} \startdata
  $e_{1i}$ &  12785986  & 6481193  & 387801 \\
  $e_{2i}$ &  -6191399  & 7506478  & 677060 \\
  $e_{3i}$ &    67818   & -312891  & 7351714 \\
\hline
  $g_{j}$   & 236$^{"}$.02   & 79$^{"}$.76  & 8$^{"}$.49
\enddata
\tablenotetext{b}{Note that $e_{13}$ and $e_{23}$ terms are
contributions from Planet D, which is much less than those of the
mutual effects of the two inner planets.}
\end{deluxetable}
\clearpage

\figcaption[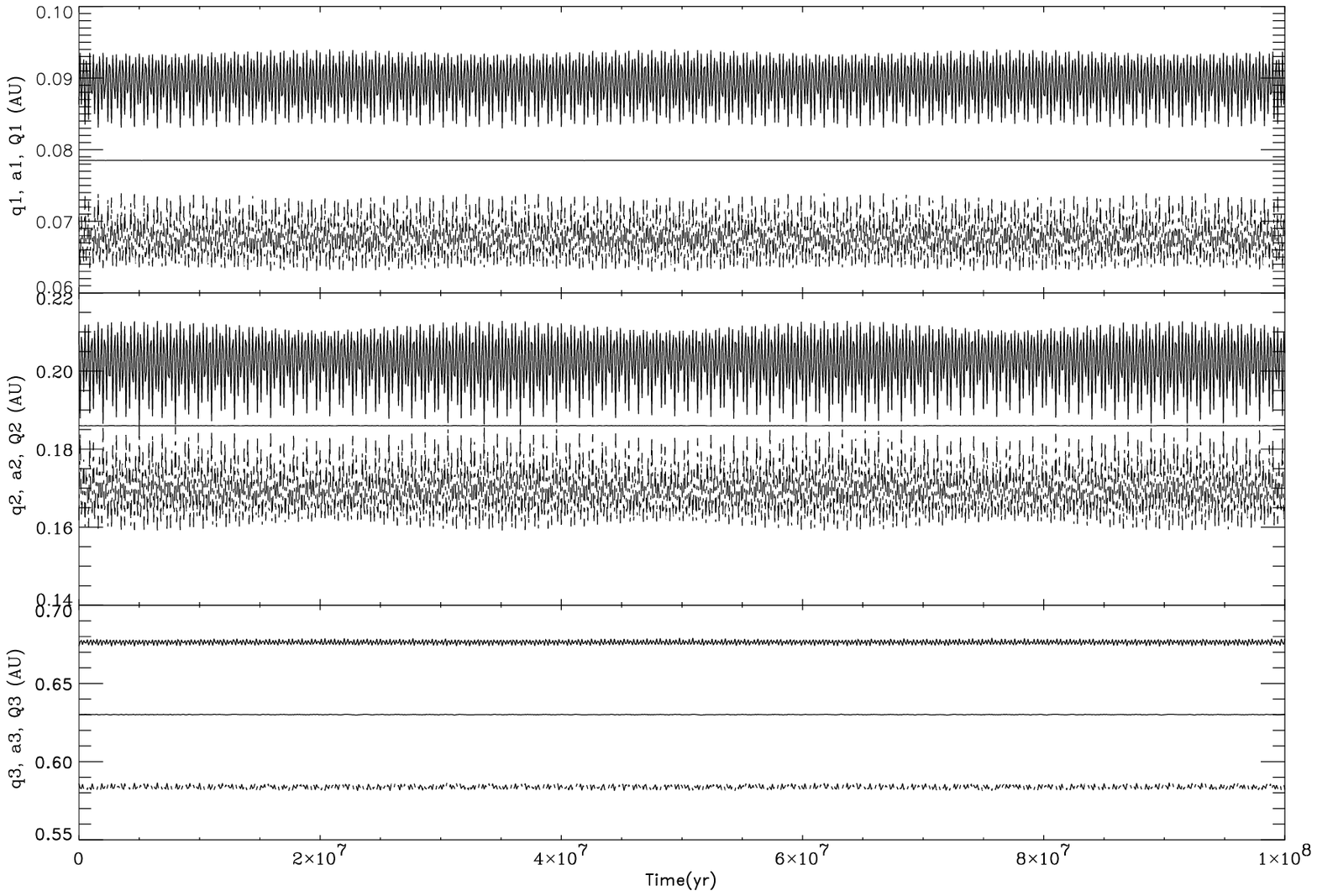]{\normalsize
Snapshot of the secular orbital
evolution of three planets is shown (for $t=10^{8}$ yr), where
$Q_{i}=a_{i}(1+e_{i})$, $q_{i}=a_{i}(1-e_{i})$ are, respectively,
the apoapsis and periapsis distances, and the straight lines are
the semi-major axes $a_{i}$. Note $a_{1}$ remains unchanged to be
0.0785 AU for 2 Gyr, and $a_{2}$ and $a_{3}$ slightly librate
about 0.186 and 0.630 AU with small amplitudes. Over the long-term
evolution, the eccentricity variations are $0.05 < e_{1}< 0.20$,
$0.0 < e_{2}< 0.15$ and $0.069 < e_{3}< 0.078$. }

\figcaption[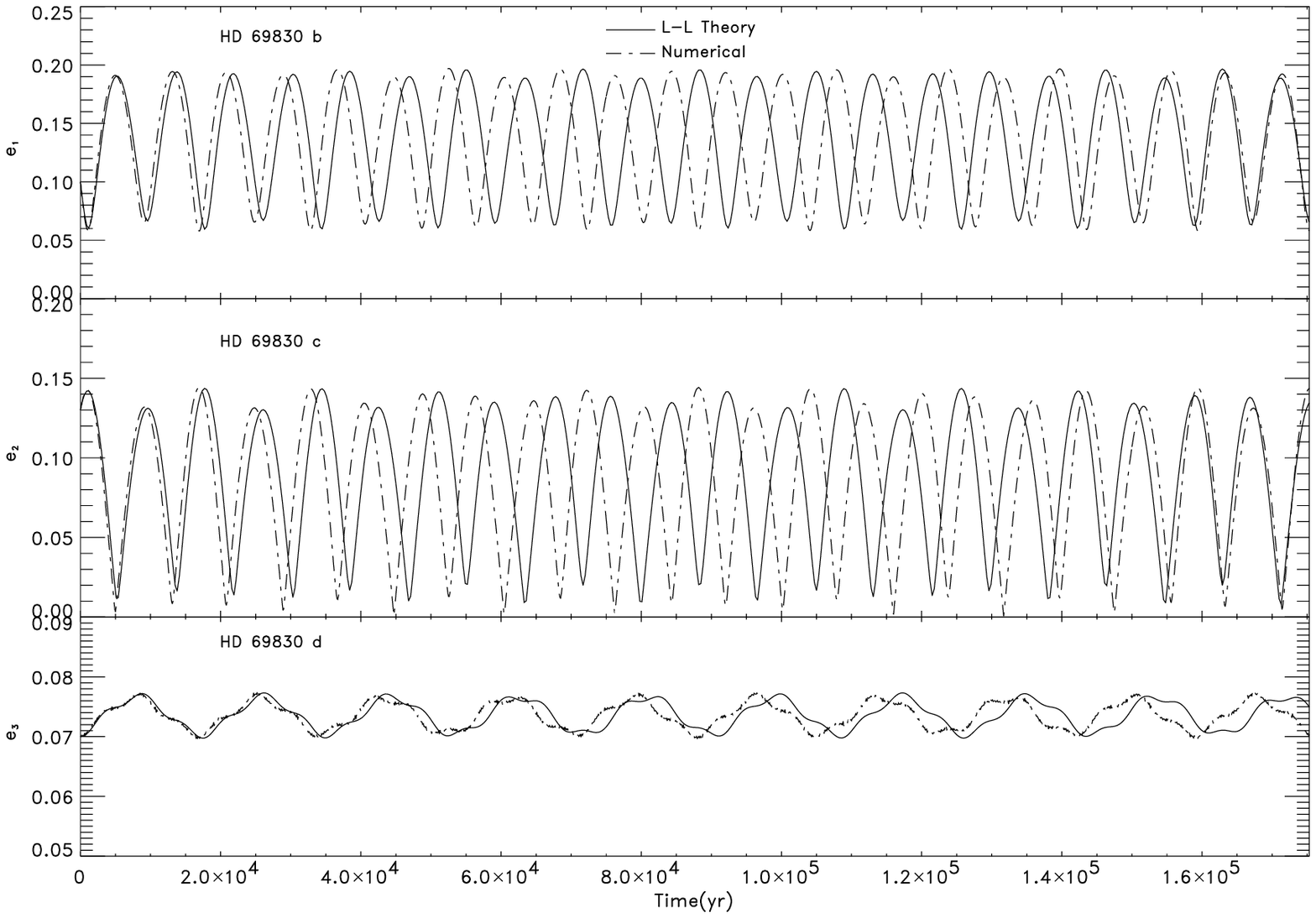]{\normalsize Comparison of secular theory with
direct numerical integration for the planetary system HD 69830.
The eccentricity variations produced by direct numerical
integration are shown by the \textit{dash dot lines}; the
corresponding eccentricity variations given by secular theory are
shown by the \textit{solid lines}. The two approaches are in good
agreement and the envelope of the variations of eccentricity are
match up extremely well.}

\figcaption[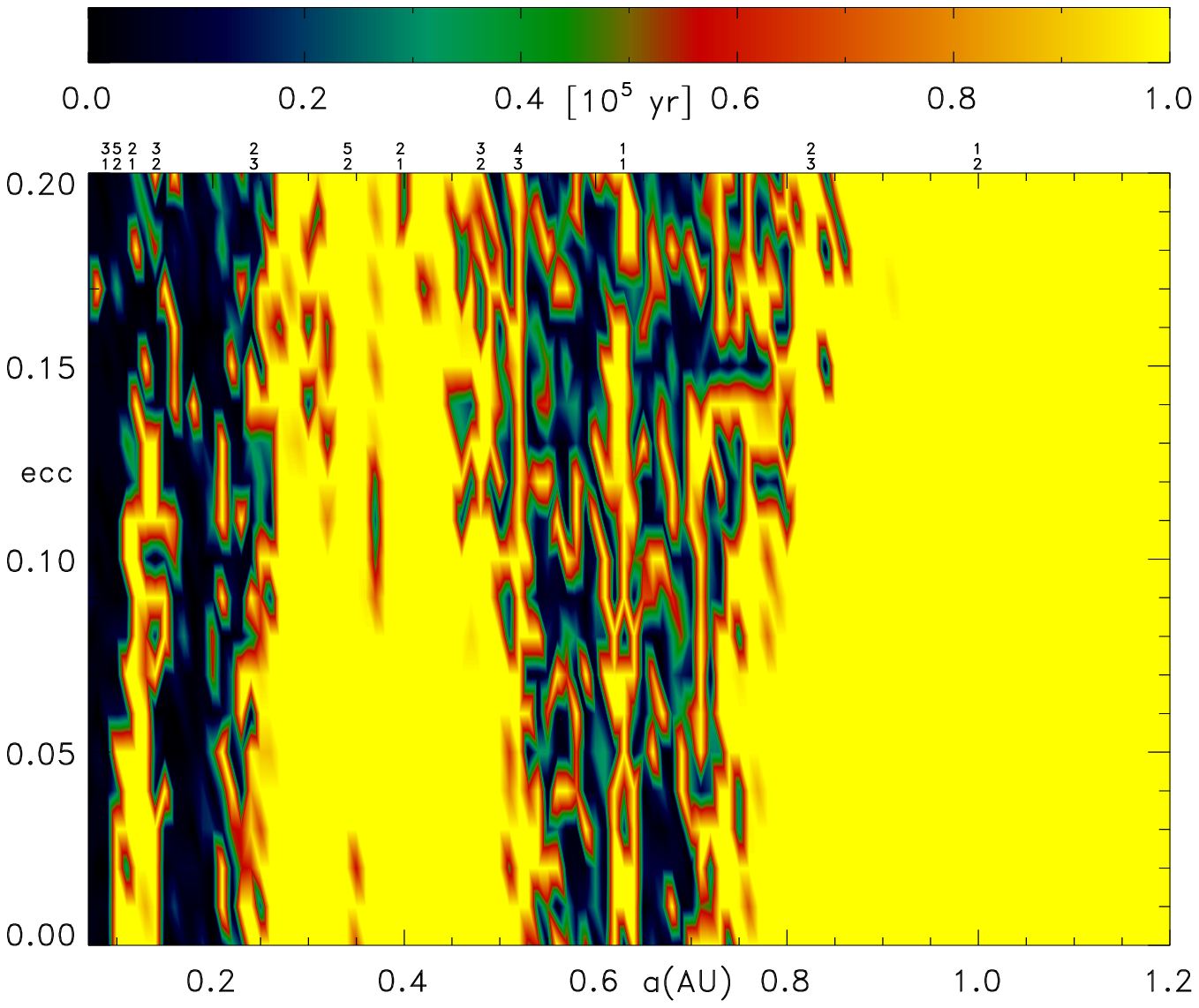, 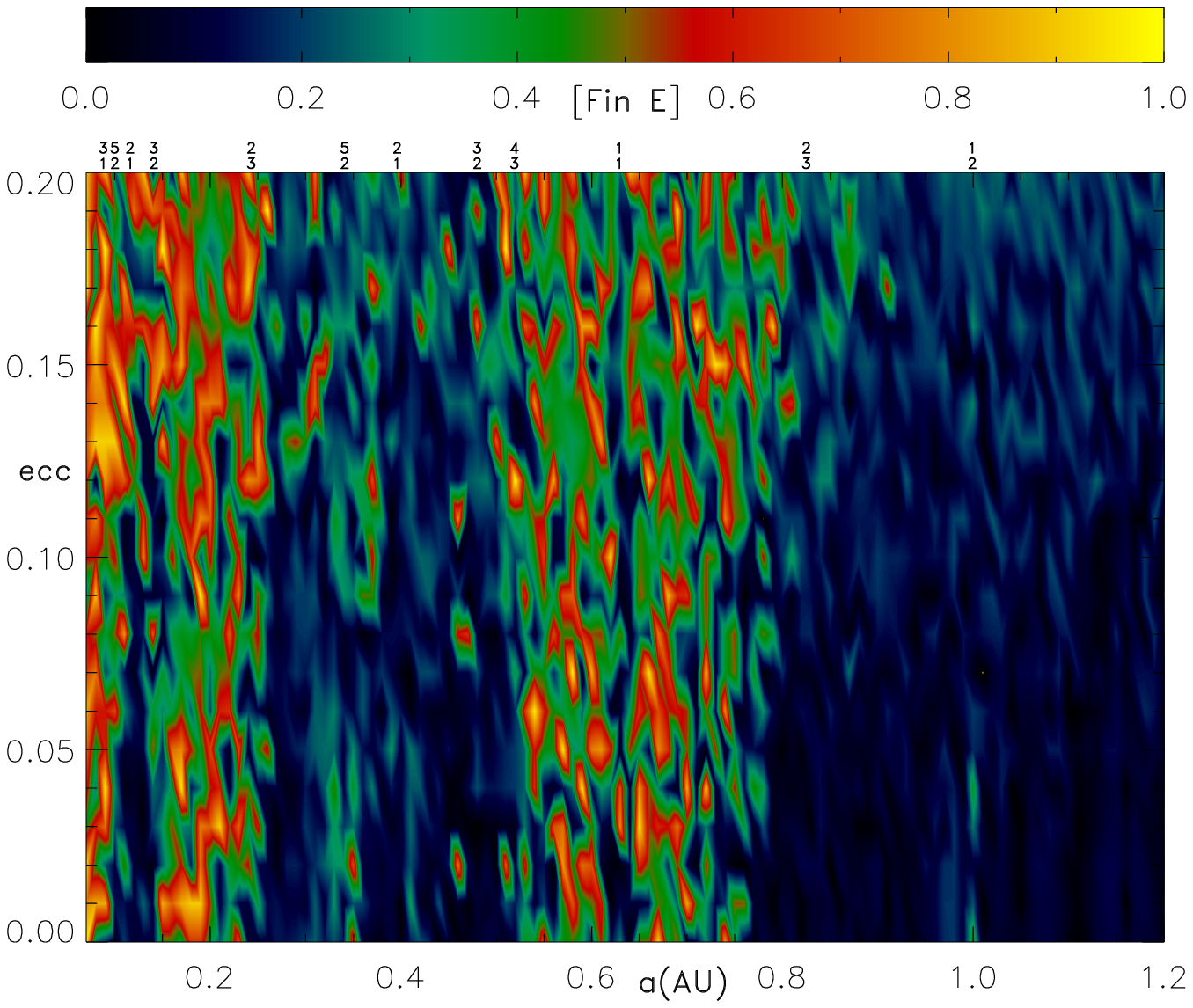]{\normalsize \textit{Left}: Contour
of the surviving time for Earth-like planets for the integration
of $10^{5}$ yr. \textit{Right}: Status of their final
eccentricities. Horizontal and vertical axes are the initial $a$
and $e$. Stable zones for the low-mass planets in the region
between 0.3 and 0.5 AU, and 0.8 and 1.2 AU with final
eccentricities of $e < 0.20$ (cool eccentricities). Unstable zones
for the nearby orbits of three planets, and the planetary embryos
there have short dynamical surviving time and the eccentricities
can be quickly pumped up to a high value $\sim 0.9$. The secular
resonance $\nu_{1}$ and $\nu_{2}$ can excite the eccentricities of
the terrestrial bodies, and the accumulation or depletion of the
asteroid belt are also shaped by orbital resonances of the outer
planets, for example, the asteroidal gaps of 2:1 and 3:2 MMRs with
the middle planet, and 5:2 and 1:2 resonances with the outmost
planet.}

\begin{figure}
 \figurenum{1}
   \plotone{f1.eps}
   \caption{}
\label{fig1}
\end{figure}

\begin{figure}
 \figurenum{2}
   \plotone{f2.eps}
   \caption{}
\label{fig2}
\end{figure}

\begin{figure}
 \figurenum{3}
   \plotone{f3a.eps}
   \caption{}
\label{fig3}
\end{figure}

\begin{figure}
 \figurenum{3}
   \plotone{f3b.eps}
   \caption{}
\label{fig3}
\end{figure}

\clearpage

\end{document}